\documentclass[pra,showpacs,twocolumn,aps]{revtex4-1}
\usepackage{slashed}
\usepackage{mathrsfs}
\usepackage{amsfonts}
\usepackage{amsmath}
\usepackage{amssymb}
\usepackage{revsymb}
\usepackage{graphicx}
\usepackage{mathrsfs}
\usepackage{bm}
\usepackage{bbm}
\usepackage{psfrag}
\usepackage{color}
\usepackage{hyperref}

\usepackage{natbib}

\usepackage[usenames,dvipsnames]{xcolor}
\usepackage{multirow,tabularx}

\newcommand{\be}{\begin{equation}} 
\newcommand{\ee}{\end{equation}} 
\newcommand{\bea}{\begin{eqnarray}} 
\newcommand{\eea}{\end{eqnarray}} 

\newcommand{\eps}{\varepsilon}
\newcommand{\mbf}[1]{\mathbf{#1}}
\newcommand{\trm}[1]{\textrm{#1}}

\newcommand{\figref}[1]{Fig. \ref{#1}}

\newcommand{\eqnref}[1]{Eq. (\ref{#1})}

\newcommand{\tabref}[1]{TABLE \ref{#1}}

\newcommand{\Ecr}{E_{\trm{cr}}}

\newcommand{\tsf}[1]{\textsf{#1}}

\newcommand{\vphi}{\varphi}

\newcommand{\nn}{\nonumber}

\newcommand{\taut}{\tilde{\tau}}
\newcommand{\e}{\mbox{e}\,}

\definecolor{oorange}{rgb}{1.0,0.5,0.0}

\newcommand{\new}[1]{#1}

\begin{document}
\title{Three-pulse photon-photon scattering}

\author{B.~King}
\affiliation{Centre for Mathematical Sciences, University of Plymouth, Plymouth, PL4 8AA, United 
Kingdom}
\email{b.king@plymouth.ac.uk}
 
\author{H.~Hu}
\affiliation{Hypervelocity Aerodynamics Institute, China Aerodynamics Research and 
Development Center, 621000 Mianyang, Sichuan, China}

\author{B.~Shen}
\affiliation{\new{State Key Laboratory of High Field Laser Physics, Shanghai Institute of Optics and Fine Mechanics, Chinese Academy of Sciences, Shanghai 201800, China  and Shanghai Normal University, Shanghai 200234, China}}

\date{\today}
\begin{abstract}
We calculate signals of real photon-photon scattering in the collision of three laser pulses. Taking goal parameters from the Station of Extreme Light at the upcoming Shanghai Coherent Light Source, we consider two scenarios: i) the collision of three optical pulses; ii) the collision of an XFEL with two optical pulses. Although experimentally more difficult to perform, we find that colliding three laser pulses offers certain advantages in the detection of scattered photons by separating their frequency, momentum and polarisation from the background.
\end{abstract}
\maketitle

\section{Introduction}
From early conceptual work in the 1930s \cite{halpern34}, to derivation of polarisation effects in constant backgrounds \cite{weisskopf36, heisenberg36} and a later reformulation in QED (quantum electrodynamics) \cite{karplus50, schwinger51}, the scattering of real photons with one another, has been a topic of much study. Recent measurements of anomalously polarised radiation emitted from magnetars \cite{mignani17}, and photon-photon scattering of high-energy virtual photons \cite{atlas17} have increased the attention given to this effect, but the observation of real photon-photon scattering has yet to be achieved. With laser-cavity set-ups such as PVLAS \cite{pvlas16} and BMV \cite{rizzo13}, experiment has inched closer to the sensitivity required by QED to permit photon-photon scattering. However, with plans to reach ever-higher intensities with focussed, pulsed laser systems \cite{danson15}, as well as dedicated upcoming experiments such as HIBEF \cite{schlenvoigt15} to measure the associated phenomenon of ``vacuum birefringence'', a first measurement and characterisation of real photon-photon scattering appears to be within sight. Not only could this represent a maturing of high-intensity laser experiments by producing a further example \cite{cole18,sarri17} of probing particle physics phenomenology, but reaching the QED level of sensitivity for photon-photon scattering will place further restraints on beyond-the-standard-model physics and dark-matter searches \cite{gies09,villalba-chavez14,villalba-chavez15,ellis17}.
\newline

The possibility of using high-intensity laser pulses to probe real photon-photon scattering, continues to be explored in the literature \cite{king15a}. In these scenarios, the choice of pump (to ``polarise'' the vacuum) is invariably an intense optical pulse, and so one way to categorise calculations is by the
choice of probe. Variations such as an optical probe \cite{lundstroem_PRL_06,king10a,hatsagortsyan11,monden11,king12,hu14b,fillion-gourdeau15,gies16,kohlfuerst18}, an X-ray probe \cite{dipiazza06,heinzl06,dinu14b,schlenvoigt15} and higher-energy probes such as synchrotron or nonlinear Compton scattered photons \cite{ilderton16, king16, homma17, meuren17b} have all been considered. Moreover, recent calculations of so-called ``tadpole'' diagrams have highlighted a further signal of photon-photon scattering that may be searched for in experiment \cite{gies17,karbstein17,schubert17}. 
\newline

In laser-based photon-photon scattering, there are three types of signals that are often calculated: i) a change in the polarisation; ii) a change in the wavevector or iii) a change in the frequency of probe photons. In the current paper, we aim to show how, when one considers the collision of \emph{three} laser pulses, each of these signals can be enhanced and combined. This work builds upon previous studies of three-beam collisions \cite{lundstroem_PRL_06, king10b, gies18} by focussing on enhancing detection signals in multiple scenarios and considering the likely combination of an XFEL probe with optical pumps. 
\newline

The paper is organised as follows. In Sec. II we outline steps taken \new{in the scattering calculation, approximations made and descriptions used for the colliding pulses}; in Sec. III we analyse the kinematics of the three-pulse collision and show \new{a wider range of scattered photon states are accessible} than in a two-pulse collision; in Sec. IV we present results for the optical and the XFEL set-ups and in Sec. V we conclude. App. A contains some results of benchmarking our two-beam and three-beam calculational method with literature results as well as assessing the accuracy of some approximations employed.

\section{Calculation of photon-photon scattering using three pulses}
The energies of photons that we will consider are of the order of those produced by XFELs (X-ray Free-Electron-Lasers) or lower, and constitute EM (electromagnetic) fields that are several orders of magnitude below the Schwinger limit of $\Ecr=m^{2}/e\approx 1.3\times 10^{16}\,\trm{Vcm}^{-1}$ (here and elsewhere, we set $\hbar=c=1$ and use $m$ and $e$ as the positron mass and charge respectively). Therefore, we will describe photon-photon scattering in the low-energy (for laser frequencies $\omega$ where $\omega/m \ll 1$), and weak-field limit (for electromagnetic field strengths $(E/\Ecr)^{2} \ll 1$), described by the interaction Lagrangian density:
\bea
 \mathcal{L} = \frac{2\alpha^{2}}{45m^{4}} \left[4\mathcal{F}^{2} + 7\mathcal{G}^{2}\right], \label{eqn:L}
\eea
where $\mathcal{F} = -F_{\mu\nu}F^{\mu\nu}/4 = (E^{2}-B^{2})/2$, $\mathcal{G} = -F_{\mu\nu}\widetilde{F}^{\mu\nu}/4 = \mbf{E}\cdot \mbf{B}$ for Faraday field tensor $F^{\mu\nu}$ and its dual $\widetilde{F}^{\mu\nu} = \eps^{\mu\nu\alpha\beta}F_{\alpha\beta}/2$, and electric and magnetic fields $\mbf{E}$ and $\mbf{B}$ \cite{jackson99}. (The applicability of the low-energy limit to laser pulse interactions has been explored in the literature \cite{dinu14a}.)
\newline

\begin{figure}[h!!]
\centering
\includegraphics[draft=false, width=0.7\linewidth]{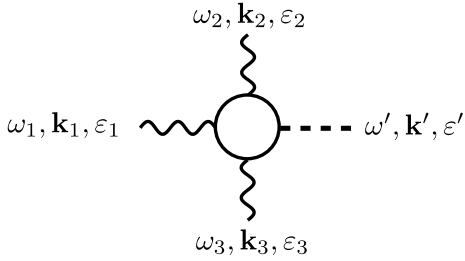}
\caption{Labelled Feynman diagram for the scattering of three laser photons (wavy lines) and the production of a signal photon (dashed line).} \label{fig:fdiag1}
\end{figure}
In this calculation, we will consider the collision of photons from three different sources to produce photons in a different outgoing state, \new{as represented by the Feynman diagram in \figref{fig:fdiag1}}. Therefore, we separate the field tensor as:
\bea
F = F_{1} + F_{2} + F_{3} + F',
\eea
where a prime ($'$) will indicate quantities in the outgoing photon state to be measured. \new{The outgoing photon will also be referred to as the \emph{signal} photon.} In some set-ups we consider, one laser pulse can be considered the ``probe'' and we choose $F_{1}$ to refer to this field. Upon insertion into the interaction, \eqnref{eqn:L}, many channels of photon-photon scattering between the three sources become apparent. We only consider those that are linear in the scattered field $F'$. This leads us to split the interaction Lagrangian as:
\[
 \mathcal{L} = \mathcal{L}_{123\,\prime} + \mathcal{L}_{122\,\prime} + \mathcal{L}_{133\,\prime} + \mathcal{L}_{112\,\prime} + \ldots,
\]
where the subscripts indicate which fields are included in the four possible positions. The total scattering matrix $\tsf{S}$ is given by:
\bea
\tsf{S} = -i \int d^{4}x~\mathcal{L} = \tsf{S}_{123\,\prime} + \tsf{S}_{122\,\prime} + \tsf{S}_{133\,\prime} + \tsf{S}_{112\,\prime} + \ldots \nn \\\label{eqn:S1}
\eea
For the laser pulses, we assume only weak focussing, such that the leading-order ``paraxial'' Gaussian pulse is a good representation of experiment, and only lowest-order terms in the focussing parameter $|\mbf{k^{\perp}}_{1,2}|/\omega \ll 1$ ($\mbf{k^{\perp}}_{1,2}$ contain the part of the wavevector perpendicular to the pulse propagation direction \cite{waters17} and $\omega$ is the pulse's central frequency) and assume the pulse is ``long'' in that the pulse envelope's phase duration satisfies $(\omega\tau)^{-1} \ll 1$ ($\tau$ is the pulse duration). Then for the pulses:
\[
 \new{F^{\mu\nu}_{s} = \partial^{\mu}A^{\nu}_{s}- \partial^{\nu}A^{\mu}_{s}, \quad s \in \{1,2,3\}}
\]
where:
\[
 \new{\partial^{\mu}A^{\nu}_{s} = \sum_{j=1}^{2} \partial^{\mu}A^{\nu}_{s}(\delta_{js})}
\]
\[
 \new{\partial^{\mu}A^{\nu}_{s}(\delta_{js}) = E_{s}\frac{k^{\mu}_{s}\eps^{\nu}_{s}}{2\omega_{s}}\frac{\mbox{e}^{-\frac{(x^{\perp})^{2}}{w^{2}_{s}} - \frac{\vphi^{2}}{\Phi^{2}_{s}}}}{\sqrt{1+\varsigma^{2}_{s}}}\mbox{e}^{i\delta_{js}\left(\vphi + \tan^{-1}\varsigma_{s} -\frac{(x^{\perp})^{2}\varsigma_{s}}{w^{2}_{s}}\right)}}
\]
and $\delta_{1s}=1$, $\delta_{2s}=-1$ is used to refer to the positive and negative frequency parts of the pulse. Various pulse quantites are defined in \tabref{tab:pulsedefinitions}. Subscripts on pulse quantities will be used to indicate which pulse they refer to.
\begin{table}
  \caption{Definitions of commonly-used symbols} \label{tab:pulsedefinitions}
  \begin{tabularx}{8cm}{p{3cm} p{5cm}}
    \hline\hline
    $E$ & peak electromagnetic field strength\\
    $\vphi = k\cdot x$ & phase \\
    $\Phi = \omega \tau$ & phase pulse duration \\  
    $w_{s,0}$ & pulse waist of pulse $s$ \\
    $w_{s} = w_{s,0}\sqrt{1+\varsigma^{2}}$ & pulse width of pulse $s$\\
    $\eps^{\mu}$ & four-polarisation \\
    $l = \omega w_{0}^{2}/2$ & Rayleigh length\\
    $\varsigma = \mbf{k}\cdot \mbf{x} / \omega l$ & curvature parameter \\
    $\mbf{n}$ & normalised wavevector $\mbf{k}=\omega \mbf{n}$\\
    \hline\hline
  \end{tabularx}
\end{table}
In addition, we assume the polarisation four-vector $\eps$ can take one of two transverse states (i.e. satisfying $k\cdot \eps = 0$), is normalised so that $\eps\cdot \eps = -1$ and is entirely spacelike in the lab frame such that $\eps^{\mu} = (0, \pmb{\eps})^{\mu}$ with $\pmb{\eps}\cdot\pmb{\eps} = 1$. The emitted photon is assumed to be in a plane wave state so that:
\bea
\partial^{\mu} A'^{\nu} = \eps'^{\nu} k'^{\mu} \frac{\e^{ik'\cdot x}}{\sqrt{2Vk'^{0}}},
\eea
where $V$ is a volumetric normalisation factor.
\newline

\new{We note that the diffraction parameter $\xi = w_{s,0}^{2}/\lambda_{s}r$ \cite{king10b} for detector distance $r$, is expected to fulfill $\xi \ll 1$, corresponding to the Fraunhofer or ``plane-wave'' limit, justifying our use of a scattering matrix approach.}
\newline

As each of the three pulses has positive and negative frequency parts, which lead to different kinematics, it will be useful to split each scattering term up into these parts. For example:
\bea
\tsf{S}_{123\,\prime}= \sum_{j_{1}=1}^{2}\sum_{j_{2}=1}^{2}\sum_{j_{3}=1}^{2} \tsf{S}_{123\,\prime}(\delta_{j_{1}},\delta_{j_{2}},\delta_{j_{3}}),
\eea
where the summand is the term linear in $A_{1}(\delta_{j_{1}})A_{2}(\delta_{j_{2}})A_{3}(\delta_{j_{3}})$. Then a subchannel to photon-photon scattering can be written as $\tsf{S}_{123'}(\delta_{j_{1}},\delta_{j_{2}},\delta_{j_{3}}) = \mathcal{I}$, where:
\bea
\mathcal{I} &=& -i C_{j_{1},j_{2},j_{3}}\int d^{4} x \,\e^{-\frac{t^{2}}{\taut^{2}}+i \widetilde{\vphi}-x^{2}\left(\mathcal{C}_{xx}+\mbf{\overline{k}_{x}}\cdot\mbf{\overline{k}_{x}}\right)-\mathcal{C}_{xy}xy}\nn \\
&& \times \e^{-y^{2}\left(\mathcal{C}_{yy}+\mbf{\overline{k}_{y}}\cdot\mbf{\overline{k}_{y}}\right)
 -z^{2}\left(\mathcal{C}_{zz}+\mbf{\overline{k}_{z}}\cdot\mbf{\overline{k}_{z}}\right)-\mathcal{C}_{xz}xz-\mathcal{C}_{yz}yz}\nn \\
&& \new{\times \e^{\taut^{2}\left(x \,\pmb{\overline{\omega}}\cdot\mbf{\overline{k}_{x}}+y \,\pmb{\overline{\omega}}\cdot\mbf{\overline{k}_{y}}+z \,\pmb{\overline{\omega}}\cdot\mbf{\overline{k}_{z}}\right)^{2}+i \widetilde{\Phi}\left(x \,\pmb{\overline{\omega}}\cdot\mbf{\overline{k}_{x}}+y \,\pmb{\overline{\omega}}\cdot\mbf{\overline{k}_{y}}+z \,\pmb{\overline{\omega}}\cdot\mbf{\overline{k}_{z}}\right)}}\nn \\
 && \times \prod_{s=1}^{3}\frac{1}{\sqrt{1+\varsigma_{s}^{2}}}\,\e^{i\delta_{j_{s}}\left[-\varsigma_{s} \frac{(x^{\perp}_{s})^{2}}{w_{s}^{2}}+\tan^{-1}\varsigma_{s}\right]} \label{eqn:I}
\eea
where $\widetilde{\vphi} = \widetilde{k} \cdot x $, \new{$\widetilde{\Phi} = \tilde{\omega}\tau$} and
\bea
 \widetilde{k} &=& \delta_{j_{1}}k_{1}+\delta_{j_{2}}k_{2}+\delta_{j_{3}}k_{3}+k' \nn \\
 \taut &=& \left(\frac{1}{\tau_{1}^{2}}+\frac{1}{\tau_{2}^{2}}+\frac{1}{\tau_{3}^{2}}\right)^{-\frac{1}{2}} \nn \\
 \overline{\omega} &=& \new{\left(\frac{\omega_{1}}{\Phi_{1}},\frac{\omega_{2} }{\Phi_{2}}, \frac{\omega_{3}}{\Phi_{3}}\right),} \nn\\
 \overline{k}_{i} &=& \left(\frac{k_{1,i}}{\Phi_{1}},\frac{k_{2,i} }{\Phi_{2}}, \frac{k_{3,i}}{\Phi_{3}}\right), \quad i\in\{x,y,z\}\nn\\
 \left(x_{l}^{\perp}\right)^{2} &=& \left(\pmb{\eps}_{l}\cdot \mbf{x}\right)^{2} + \left(\mbf{n}_{l}\wedge\pmb{\eps}_{l}\cdot \mbf{x}\right)^{2}, \quad l \in \{1,2,3\} \nn \\
 \left(x_{l}^{\perp}\right)^{2} &=& C_{l,xx}x^{2}+C_{l,yy}y^{2}+C_{l,zz}z^{2}\nn \\
 && +C_{l,xy}xy+C_{l,xz}xz+C_{l,yz}yz \nn \\
 \mathcal{C}_{ij} &=& \frac{C_{1,ij}}{w_{1}^{2}}+\frac{C_{2,ij}}{w_{2}^{2}}+\frac{C_{3,ij}}{w_{3}^{2}}\quad i,j\in\{x,y,z\}\nn\\
\eea

We note that the structure of \new{other scattering channels with three pulses, e.g.} $\tsf{S}_{122'}$ of $\tsf{S}_{133'}$, is analogous to \eqnref{eqn:I}. The integral in \eqnref{eqn:I} does not, in general, permit a closed form solution due to the presence of the curvature terms. However, if uses the \new{\emph{infinite Rayleigh-length approximation} (IRLA)}, i.e. assumes $l \to \infty$, \eqnref{eqn:I} just becomes the integral of a four dimensional Gaussian with oscillatory phase. From previous work \cite{king12}, we expect the IRLA to be a good approximation when $l \gg \trm{min}\left\{\tau,L^{\parallel}\right\}$, where $L^{\parallel}$ is the dimension of the effective interaction region of the three beams. We then find:
\bea
\mathcal{I} &=& -i C_{j_{1},j_{2},j_{3}}\pi^{3/2}XYZ\e^{-\left[\frac{\tilde{\omega}\tilde{\tau}}{2}\right]^{2}-\left[\frac{\Omega_{x}X}{2}\right]^{2}-\left[\frac{\Omega_{y}Y}{2}\right]^{2}-\left[\frac{\Omega_{z}Z}{2}\right]^{2}}\nn \\
\eea
where 
\bea
 X^{-2} &=& \mathcal{C}_{xx}+\mbf{\overline{k}_{x}}\cdot\mbf{\overline{k}_{x}}-\left(\taut~\pmb{\overline{\omega}}\cdot\mbf{\overline{k}_{x}}\right)^{2} + Z^{2}\mathscr{C}_{xz}^{2}
 \nn \\ && +Y^{2} \left[\mathscr{C}_{xy}+Z^{2}\mathscr{C}_{xz}\mathscr{C}_{yz}\right]^{2} \nn \\
 Y^{-2} &=& \mathcal{C}_{yy}+\mbf{\overline{k}_{y}}\cdot\mbf{\overline{k}_{y}} -\left(\taut~\pmb{\overline{\omega}}\cdot\mbf{\overline{k}_{y}}\right)^{2} + Z^{2}\mathscr{C}_{yz}^{2} \nn \\  
 Z^{-2} &=& \mathcal{C}_{zz}+\mbf{\overline{k}_{z}}\cdot\mbf{\overline{k}_{z}}-\left(\taut~\pmb{\overline{\omega}}\cdot\mbf{\overline{k}_{z}}\right)^{2}\nn\\
 \Omega_{x} &=& -\tilde{k}_{x}+\tilde{\omega}\taut^{2}\,(\pmb{\overline{\omega}}\cdot\mbf{\overline{k}_{x}}) + \Omega_{z} Z^{2} \mathscr{C}_{xz} \nn \\ 
 && + \Omega_{y}Y^{2} \left[\mathscr{C}_{xy}+Z^{2}\mathscr{C}_{xz}\mathscr{C}_{yz}\right] \nn \\
  \Omega_{y} &=& -\tilde{k}_{y}+\tilde{\omega}\taut^{2}\,(\pmb{\overline{\omega}}\cdot\mbf{\overline{k}_{y}})+\Omega_{z} Z^{2} \mathscr{C}_{yz}\nn \\
 \Omega_{z} &=& -\tilde{k}_{z}+\tilde{\omega}\taut^{2}\,(\pmb{\overline{\omega}}\cdot\mbf{\overline{k}_{z}})\nn \\
 \mathscr{C}_{ij} &=& \taut^{2}~\pmb{\overline{\omega}}\cdot\mbf{\overline{k}_{i}}~\pmb{\overline{\omega}}\cdot\mbf{\overline{k}_{j}}-\mathcal{C}_{ij},\quad i,j\in\{x,y,z\}.
\eea
To calculate the total number of photons scattered, $\tsf{N}$, we integrate over outgoing photon momenta:
\bea
 \tsf{N} = \int \frac{d^{3}k'}{(2\pi)^{3}} |\tsf{S}|^{2}.
\eea
For total photon yields, we will sum over the two outgoing polarisation states.


\section{Kinematics}
Already we see from the integral in \eqnref{eqn:I}, the appearance of energy-momentum conservation. The main component of the oscillating phase is given by $\widetilde{\vphi}=\widetilde{k}\cdot x$, so that the integral has a maximum at the point where the total momentum, $\widetilde{k}$, satisfies $\widetilde{k}=0$. This is unsurprising: we are assuming the pulses can be well-described by the leading-order paraxial Gaussian pulse and so photons in these pulses should have a momentum close to that from a plane wave. 
\newline

\begin{figure}[h!!]
\centering
\includegraphics[width=0.7\linewidth]{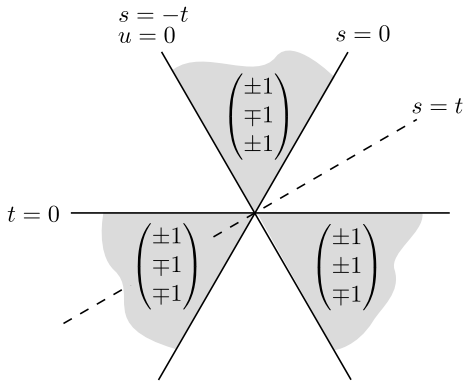}
\caption{Mandelstam plot for the permitted regions (shaded) of the kinematic invariants $s$, $t$ and $u$, where the channels for different combinations of laser photons being absorbed/emitted are indicated by column vectors $(\delta_{j_{1}},\delta_{j_{2}},\delta_{j_{3}})^{\tsf{T}}$. The permitted region for $\delta_{j_{1}}=\delta_{j_{2}}=\delta_{j_{3}}=\pm1$ is a single point at the origin.} \label{fig:mstam1}
\end{figure}

To study the kinematics of the reaction, let us write energy-momentum conservation for plane-waves:
\bea
\delta_{j_{1}}k_{1}+\delta_{j_{2}}k_{2}+\delta_{j_{3}}k_{3}+k' =0.
\eea
The solution to this equation for $k'$ allows a neat analysis using Mandelstam variables \cite{landau4}:
\bea
 s &=& (\delta_{j_{1}}k_{1} + \delta_{j_{2}}k_{2})^{2} = 2 \delta_{j_{1}}\delta_{j_{2}} k_{1}\cdot k_{2} \nn \\
 t &=& (\delta_{j_{1}}k_{1} + \delta_{j_{3}}k_{3})^{2} = 2 \delta_{j_{1}}\delta_{j_{3}} k_{1}\cdot k_{3} \nn \\
 u &=& (\delta_{j_{2}}k_{2} + \delta_{j_{3}}k_{3})^{2} = 2 \delta_{j_{2}}\delta_{j_{3}} k_{2}\cdot k_{3}. \label{eqn:mstam0}
\eea
We see that $s+t+u=0$ and by squaring $L_{\lambda} = \eps_{\lambda \alpha\beta\gamma}k_{1}^{\alpha}k_{2}^{\beta}k_{3}^{\gamma}$, that $s\,t\,u \geq 0$. Then it follows that if $u=0$ and $s=-t$ are the same line, the usual triangular region of the Mandelstam plot collapses to a point, as depicted in \figref{fig:mstam1}. As the inside of the point-sized triangular region corresponds to the case that $\delta_{j_{1}}=\delta_{j_{2}}=\delta_{j_{3}}=\pm1$, this immediately tells us that the absorption or emission of three photons and their conversion to an outgoing scattered photon is only kinematically allowed if $s=t=u=0$, or in other words, if the photons are propagating parallel. This case corresponds to the scattering of a pure plane wave, for which the electromagnetic invariants disappear and hence so does the probability for the scattering to take place. 
\newline

As we see from the various terms in the scattering matrix \eqnref{eqn:S1}, in a three-beam interaction, each interaction between two beams is also included, but they are kinematically much more restricted.

\subsection{Limitations of a two-beam set-up}
The Mandelstam plot allows us to see the kinematic limitation of having just two colliding beams. Suppose we set $k_{3}=k_{2} + \Delta k$, then $u=2\delta_{j_{2}}\delta_{j_{3}}k_{2}\cdot \Delta k$. For plane waves, $\Delta k =0$, and the accessible kinematic region shrinks to the intersection of $u=0$, $\delta_{j_{2}}s = \delta_{j_{3}} t$ and $s=-t$. Immediately we see that if $\delta_{j_{2}}\delta_{j_{3}} = 1$, then the only solution is $s=t=u=0$, which we have already ruled out as having zero probability. But this solution corresponds to the channel where two photons are absorbed to or emitted from the same pulse, which we therefore conclude as being prohibited for plane waves. This leaves only $\delta_{j_{2}}\delta_{j_{3}} = -1$, where the permitted kinematic region is just a line along the $u=0$ axis. But in this case, since now $k' = -\delta_{j_{1}}k_{1}-\left(\delta_{j_{2}}+\delta_{j_{3}}\right)k_{2}$, we must infer that $k'=k_{1}$ meaning there can be no signal of photon-photon scattering in the momentum or frequency of the measured photon, as these are fixed to that of the probe ($\delta_{j_{1}}=1$ corresponds to the negative-frequency solution and is ruled out for the emitted signal photon). 
\newline

\begin{figure}
\centering
\includegraphics[width=0.7\linewidth]{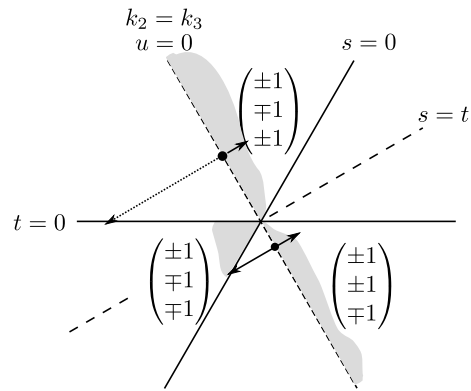}
\caption{Representative allowed kinematic regions (shaded) for a two-beam set-up using a pulsed focussed pump laser. Arrows represent in which directions the allowed kinematic region for a monochromatic set-up at fixed $s$ and $t$, can widen as a finite pulse duration and focussing effects are included. It is harder to reach the kinematic region for frequency-shifting (bottom-left region).} \label{fig:mstam2}
\end{figure}

If the pulse has finite duration, it has a finite bandwidth $\Delta \omega$, and if it is focussed, it has a finite momentum spread $\Delta\mbf{k}$. By increasing the magnitude of $k_{2}\cdot \Delta k$, one increases the magnitude of $u$ and hence the kinematic region can expand in either of the directions perpendicular to the $u=0$ line. However, we expect pulses used in experiment to be well-approximated by the paraxial conditions mentioned above: $\omega \tau \gg 1$, $|\mbf{k^{\perp}}_{1,2}| \ll \omega$, so the magnitude of the components of $\Delta k$ is limited by: $\Delta \omega / \omega \ll 1$ and $|\Delta \mbf{k}_{i}|/ \omega \ll 1$ respectively.  Therefore, only a small region of the Mandelstam plot around $u=0$ is accessible to a two-beam set-up. This situation is depicted in \figref{fig:mstam2}, where the accessible regions have been reduced to a strip around $u=0$. The arrows show two situations as $\Delta k$ is increased from zero for a specific $s$ and $t$. When $\Delta k = 0$, the situation is as for a collision of plane waves and the kinematical region is a single point on the axis. As the width of frequencies and momenta is increased from zero, depending on the mutual sign between $\delta_{j_{2}}$ and $\delta_{j_{3}}$, the region of allowed kinematics can widen, but it must overlap with the overall allowed three regions shown in \figref{fig:mstam1}. For any non-trivial situation, $s$ and $t$ will be non-zero. It can be seen that, whatever the value of $s$ and $t$, the permitted region can always expand when one photon is absorbed and one emitted from the pump ($\delta_{j_{2}}\delta_{j_{3}}=-1$). However, for two photons to be emitted to or absorbed from the pump ($\delta_{j_{2}}\delta_{j_{3}}=1$), there must be enough support from the bandwidth of the pump lasers for a large enough $\Delta k$ to reach the $\delta_{j_{2}}=\delta_{j_{3}}=-1$ frequency-shifting region. For a given $\Delta k$ this requires \new{$u(\Delta k)\geq 2t(\Delta k=0)$}, or equivalently \new{$k_{2}\cdot \Delta k \geq 2 k_{1}\cdot k_{2}$}. Therefore, for higher probe frequencies, it requires a larger bandwidth from the pump lasers to reach two photon emission. 
Thus under these conditions, two-beam set-ups only permit \new{absorption accompanied by emission (and vice versa)} of a pump photon. This has the consequence, for our example, that the scattered photon wavevector is only displaced from the probe by $k' - (-\delta_{j_{1}}k_{1}) = \delta_{j_{3}} \Delta k$, i.e. by the frequency and wavevector bandwidth of the pump laser.

\section{Three-beam scattering}
In this section, we show how a three-beam set-up can be used to deliver clearer experimental signatures of photon-photon scattering than two-beam scenarios. We consider two possibilities: i) the collision of three optical pulses; ii) the collision of an X-ray probe with two optical pump beams. In these examples, we use the goal parameters from the Station of Extreme Light \new{\cite{shen18}} at the Shanghai Coherent Light Source, where there are plans to construct four combinable $25\,\trm{PW}$ optical pulsed lasers at $1.365\,\trm{eV}$ ($910\,\trm{nm}$) \new{and the Shanghai XFEL, which will provide $10^{12}$ X-ray photons at $3-15\,\trm{keV}$}. 
\newline

For the scattered photons, we will choose one polarisation state $\eps'^{\parallel} = (0,\pmb{\eps'^{\parallel}})$ to be that polarisation parallel to the polarisation of the probe if the central probe wavevector is rotated into the scattered photon wavevector (without rotation around the probe propagation axis), and $\eps'^{\perp} = (0,\mbf{n'}\wedge\pmb{\eps'^{\parallel}})$, where $\mbf{k'} = \mbf{n'} \omega'$.

\subsection{Optical set-up}
\begin{figure}[h!!]
\centering
\includegraphics[width=0.7\linewidth]{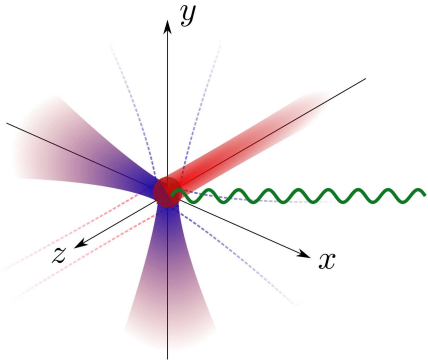}
\caption{Optical three-beam set-up with signal photons depicted in green.} \label{fig:setup1}
\end{figure}
\begin{figure}[h!!]
\centering
\includegraphics[width=0.7\linewidth]{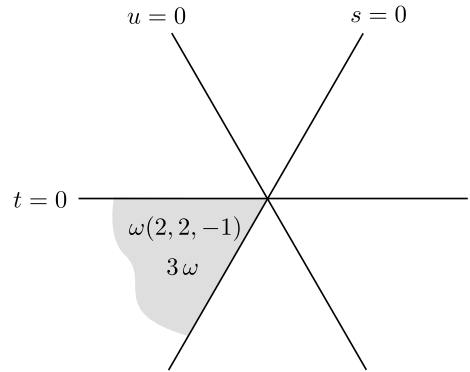}
\caption{Mandelstam plot with frequency and momentum labels for the three-beam optical set-up suggested in \cite{lundstroem_PRL_06}.} \label{fig:mstam3}
\end{figure}

The largest frequency shift in the scattered photons relative to the lasers used to generate those photons, can be attained by using laser pulses with similar frequncies. For this, an optical set-up is ideal. Specifically, we take the suggestion from the work by Lundstr\"om and colleagues \cite{lundstroem_PRL_06}, but include in the description of the laser pulses both pulse duration and focussing terms. The inspired suggestion in \cite{lundstroem_PRL_06} is to have:
\bea
 k_{1} &=& \omega\{1,0,0,1\} \nn \\
 k_{2} &=& 2\omega\{1,1,0,0\} \nn \\
 k_{3} &=& 2\omega\{1,0,1,0\} \nn,
\eea
which gives $k' = -\omega\{\delta_{j_{1}}+2(\delta_{j_{2}}+\delta_{j_{3}}),2\delta_{j_{2}},2\delta_{j_{3}},\delta_{j_{1}}\}$. A schematic of the optical set-up is shown in \figref{fig:setup1} and the corresponding Mandelstam plot for the three-beam interaction, with frequency and normalised wavevector labels, is given in \figref{fig:mstam3}. We see immediately, unlike a typical two-beam set-up, the frequency-shifting sector is accessible and generates a signal spatially separated from the laser propagation axes. 
\newline

\begin{figure}[h!!]
\centering
\includegraphics[width=0.7\linewidth]{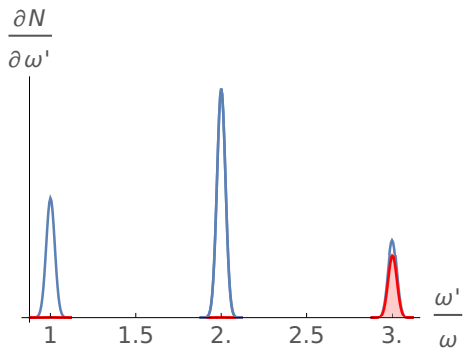}
\caption{Spectrum of scattered photons in the optical set-up. Fundamental and second harmonics are from two-beam scattering. Filled red curves indicate \new{the proportion} photons scattered into a different polarisation state to the ``probe''.}\label{fig:3beamFreq}
\end{figure}

\begin{figure}[h!!]
\centering
\includegraphics[width=0.7\linewidth]{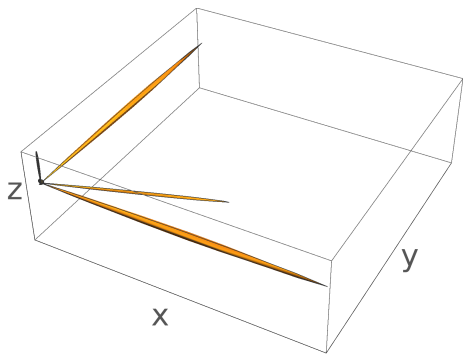}
\caption{The angular distribution of scattered photons $\partial N/\partial \Omega$. The photons scattered almost parallel to the co-ordinate axes, are those generated in two-beam interactions.}\label{fig:3beamSph}
\end{figure}

As an example, for the optical set-up we use three $25\,\trm{PW}$ pulses of duration $30\,\trm{fs}$, and focal width of $5\,\mu\trm{m}$ to generate  the frequency spectrum in \figref{fig:3beamFreq} and the angular spectrum in \figref{fig:3beamSph}. A total of $6400$ photons were scattered, the results of which are summarised in \tabref{tab:3beamoptical}. Scattered photons with frequencies equal to the fundamental and second harmonic, propagate almost parallel to the pulses that created them, demonstrating in a clearer way how three-beam scattering includes two-beam scattering within the kinematics. Photons scattered into the third harmonic propagate in a clearly different direction to the three pulses used to generate them. 
\begin{center}
\begin{table}
  \caption{Three-beam optical set-up results} \label{tab:3beamoptical}
  \begin{tabularx}{6cm}{c r rl}
    \hline\hline
Frequency~~ & $\eps'^{\parallel}$ photons~~  & \multicolumn{2}{c}{\hspace{0.2cm}$\eps'^{\perp}$ photons}\\
$\omega$~~ & \hspace{0.5cm}$1180$~~& $0$&\!\!$.7$ \\
$2\,\omega$~~ & $4460$~~& $0$&\!\!$.7$ \\
$3\,\omega$~~ & $150$~~& \hspace{0.5cm}$610$&  \\
    \hline\hline
  \end{tabularx}
\end{table} 
\end{center}
In \tabref{tab:3beamoptical}, the scattered photon polarisation $\eps'^{\parallel}$ is parallel to the pulse that acted as the ``probe'' in the scattering channel. For example, $\eps'^{\parallel}$ for the $\tsf{S}_{112'}$ term is parallel to $\eps_{2}$. From \tabref{tab:3beamoptical}, we see that for the optical set-up, two-beam scattering does not give a strong signal in the polarisation of the scattered photons (less than $0.1\,\%$), whereas in the three-beam set-up, polarisation can be used as an extra signature of photons that have been scattered (about $80\,\%$ of them).


\subsection{XFEL + Optical set-up}
Let us choose the X-ray probe to be pulse $1$, and the two optical pumps to be pulses $2$ and $3$. Then from the Mandelstam variables defined in \eqnref{eqn:mstam0}, we see that, unless one of the optical pumps is propagating almost parallel to the X-ray photons, $u \ll s,t$. In particular, \new{$u(\Delta k) < 2t(\Delta k = 0)$}, and so the frequency-shifting sector open to the optical set-up in \figref{fig:mstam3} is \emph{not} accessible in the XFEL case. However, if one assumes a highly collimated beam and a narrow bandwidth for the XFEL, we will see that by adjusting the optical pulses, it is still possible to generate a signal of photon-photon scattering in the frequency and wavevector of measured photons using a three-beam interaction.
\newline

Since the XFEL is much weaker than the two optical pump pulses, one can make the approximation that:
\[
 \tsf{S} \approx \tsf{S}_{123\,\prime}+ \tsf{S}_{122\,\prime} + \tsf{S}_{133\,\prime},
\]
i.e. that the pump pulses are, to a good approximation, not scattered by the XFEL.

\subsubsection{Orthogonal set-up}
\begin{figure}[h!!]
\centering
\includegraphics[width=0.7\linewidth]{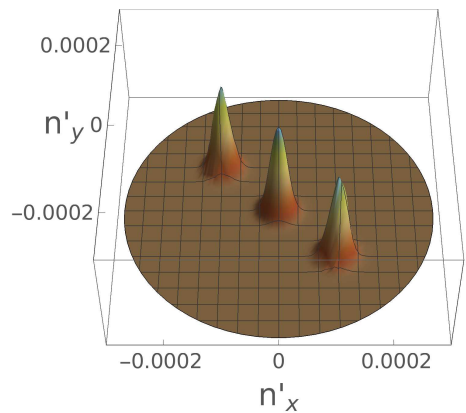}
\caption{The angular distribution of scattered photons $\partial N/\partial \Omega$, where $n'_{x}$ and $n'_{y}$ are the transverse components of the normalised wavevector $\mbf{n'}$ where $\mbf{k}' = \mbf{n'}\omega'$.}\label{fig:XrayAngles1}
\end{figure}

The largest spatial separation of the probe background from the scattered photons, will be achieved if the momentum change is concentrated \emph{orthogonal} to the probe propagation direction. For this reason, we consider an analogous set-up to the all-optical case, but now with an XFEL probe:
\bea
 k_{1} &=& \omega_{\tsf{x}}\{1,0,0,1\} \nn \\
 k_{2} &=& \omega\{1,1,0,0\} \nn \\
 k_{3} &=& \omega\{1,0,1,0\} \nn,
\eea
giving $k' = -\{\omega_{\tsf{x}}\delta_{j_{1}}+(\delta_{j_{2}}+\delta_{j_{3}})\omega,\delta_{j_{2}}\omega,\delta_{j_{3}}\omega,\delta_{j_{1}}\omega_{\tsf{x}}\}$. Immediately we see that $\delta_{j_{1}}=-1$ to ensure photons are scattered with positive energy, and therefore, the signal will propagate almost parallel to the probe. Also, we see that the typical polar angle of the scattered photons (defined with respect to the probe propagation axis) will be of the order of $\omega/\omega_{\tsf{x}}$. Using $12.9\,\trm{keV}$ as the central X-ray photon energy in a pulse of duration $10\,\trm{fs}$, focussed to $1\,\mu\trm{m}$ and collided with two $50\,\trm{PW}$ laser pulses of duration $20\,\trm{fs}$, focussed to $5\,\mu\trm{m}$ each, we find the angular spectrum in \figref{fig:XrayAngles1}.
\newline

The numbers of photons scattered into the central peak in \figref{fig:XrayAngles1} was $80$, and the total number scattered into the side peaks, which are due to the three-beam interaction, was $170$. The side peaks are at an angle of around $0.16\,\trm{mrad}$, so on a detector $5\,\trm{m}$ away from the interaction, this is a spatial separation of around $0.8\,\trm{mm}$. Moreover, all of the scattered photons have polarisation $\eps'^{\perp}$ and are hence distinguishable from the probe photons. However, this set-up of pulses has only a very low change in energy of the scattered photons, of the order of $1\,\trm{meV}$. 
\newline

\begin{figure}[h!!]
\centering
\includegraphics[width=0.7\linewidth]{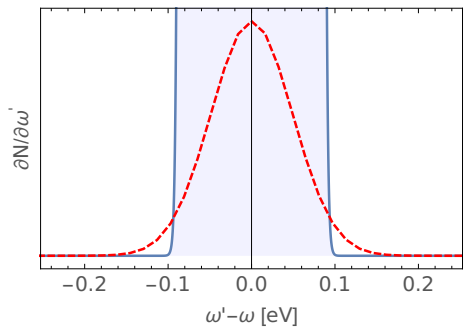}
\caption{\new{Orthogonal set-up: scattered photons are shown by the red dashed line have a larger width than the X-ray background photons shown by the filled blue curve.}
}\label{fig:xrayenergyperpshowcase}
\end{figure}

\new{To increase the energy difference of the scattered photon, one might consider using very short optical pulses, which have a wider bandwidth than the X-ray, so that the spread of scattered photons' energies would be sufficient to distinguish them from the probe background. For example, if a long X-ray pulse is chosen at $50\,\trm{fs}$, focussed to $0.1\,\mu\trm{m}$, and if the optical pulses are chosen short at just $10\,\trm{fs}$, focussed to $5\,\mu\trm{m}$, then the scattered photons have the spectrum in \figref{fig:xrayenergyperpshowcase}}.
\new{$8$ of the $170$ total scattered photons are at energies and also polarisations that are distinct from the probe background. However, these scattered photons all originate from the two-pulse interactions, and it is clear that the energy resolution of the detector would play a large role in the measurement of these photons.}

\subsubsection{``\tsf{y}'' set-up}
Since it is kinematically unfavourable to absorb or emit two pump photons, the difference in energy of the scattered photons from the probe background must come from the wider bandwidth of the optical pulses. This is clear simply from energy conservation: {\protect $\omega' = \omega_{\tsf{x}}+\delta_{j_{2}}(\omega_{2}-\omega_{3})$}. If we set $\omega_{3}=\omega_{2}+\Delta\omega$, where $\Delta \omega$ represents the difference in energy that can be acquired by absorbing and emitting photons from different parts of the spectrum of the pump pulses, then since $\omega_{2}/\omega_{\tsf{x}}\ll1$, scattered photons obey the vacuum dispersion relation $k'^{2}=0$ (and hence propagate to the detector) if:
\bea
\cos\theta_{3} - \cos\theta_{2} = \frac{\Delta \omega}{\omega_{2}}~(1-\cos\theta_{2}) \label{eqn:cos3}
\eea
where $\theta_{2}$ and $\theta_{3}$ are the angles between the pulses and the probe. \eqnref{eqn:cos3} shows two important points about the energy of the scattered photons: i) if we have only two pulses, i.e. if $\theta_{2}=\theta_{3}$ and if $\theta_{2} \neq \pi$, then $\Delta\omega/\omega_{2} \approx 0$; ii) we cannot just choose any relative angle between the beams and expect a signal in the energy, because the scattered photons only have a finite support from the bandwidth of the pulses.
\newline

\begin{figure}[h!!]
\centering
\includegraphics[width=0.7\linewidth]{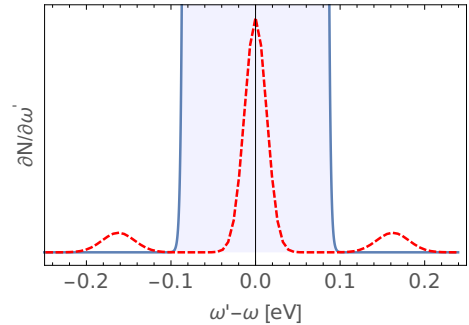}
\caption{Scattered photons are shown by the red dashed line and the filled blue curve is the spectrum of the probe background.}
\label{fig:xrayenergyshowcase}
\end{figure}

For these reasons, to maximise the energy shift of the scattered photons, we choose a set-up in which one pump pulse collides head-on with the probe, and the other pump pulse collides at an angle, making the colliding beams resemble a ``$\tsf{y}$''. We consider $12.9\,\trm{keV}$ as the central X-ray photon energy in a pulse of duration $50\,\trm{fs}$, focussed to $1\,\mu\trm{m}$ and collided with two $50\,\trm{PW}$ laser pulses of duration $10\,\trm{fs}$, focussed to $5\,\mu\trm{m}$ each. Then if we take the bandwidth of each optical pulse to be $1/\omega\tau\approx 0.05$, so that $\Delta \omega / \omega_{2} \approx 0.1$ and choose $\theta_{2}=\pi$,  \eqnref{eqn:cos3}, gives a maxmimum change in $\theta_{3}$ of around $0.2\,\pi$. Therefore, we take $\mbf{k}_{2} = \omega_{2}(0, 0, -1)$ and $\mbf{k}_{3} = \omega_{3}(\sin\theta_{3},0,\cos\theta_{3})$, with $\theta_{3} = 0.8\pi$ and find the energy spectrum in \figref{fig:xrayenergyshowcase}.
\newline

    In \figref{fig:xrayenergyshowcase}, one immediately notices the differences between the two-beam interaction, (central peak), which supplies $230$ photons, and the two sidebands due to the three-beam interaction, which supply a total of $60$ photons. Therefore, the three-beam interaction signal is separable from the probe background, for a detector resolution of the order of $0.1\,\trm{eV}$ \new{(this should be contrasted with the two-pulse signal in \figref{fig:xrayenergyperpshowcase})}. Furthermore, this corresponds to the three-beam interaction having an angular separation of around $0.075\,\trm{mrad}$. If the XFEL is focussed to only $1\,\mu\trm{m}$, then on a detector placed $5\,\trm{m}$ away from the interaction centre, the probe pulse has a width of $0.02\,\trm{mm}$ compared to the signal's $0.075\,\trm{mm}$ position. Although this is not enough to spatially separate the scattered photons from the probe background, at least the background is significantly lower than the forward-scattered two-beam interaction photons. Moreover, when the polarisation of the three pulses is chosen the same as other examples in this paper, then the scattered photons are all scattered into the $\eps'^{\perp}$ polarisation state, and hence are distinguishable from the background probe photons.


\section{Conclusion}
We have investigated photon-photon scattering in the collision of three laser pulses. This builds on the results of an initial suggestion by Lundstr\"om et al. \cite{lundstroem_PRL_06} of colliding three optical beams propagating orthogonally to one another. In particular, we have: i) introduced pulse durations and hence a finite bandwidth of the colliding pulses; ii) included focussing terms and hence a finite momentum spread of photons colliding from each pulse and iii) considered also an ``XFEL + two optical pulses'' set-up. In both the optical and XFEL set-ups, we found a common theme that the three-pulse interaction offered a signal of photon-photon scattering in: i) a frequency shift; ii) a momentum shift; iii) a flipped polarisation. Moreover, the role of pulse duration and focussing effects was crucial in generating signals separable from the background in the XFEL set-up. Although the collision of three pulses is experimentally more challenging to perform than a two-pulse collision, by using all three signals provided in a three-pulse set-up, the advantages in detection may ultimately make such a set-up preferable for performing the first ever measurement of real photon-photon scattering.

\section{Acknowledgments}
B. K. acknowledges funding from Grant No. EP/P005217/1 and H. Hu acknowledges the National Natural Science Foundation of China under Grant No. 11774415. \new{This work is supported by the Project of Shanghai Coherent Light Facility (SCLF).}

\appendix

\section{Benchmarking}
In order to have confidence in our results, we perform three types of benchmarking: i) comparison of scattering rates for the two-pulse case, with expected behaviour from literature results; ii) comparison of photon yields with the three-beam calculation of \cite{lundstroem_PRL_06}; iii) assessment of the accuracy of the IRLA by comparison with fully numerically-integrated results.

\subsection{Behaviour of scattering rates for two-pulse case}
To determine the expected behaviour in the two-pulse case, we compare our scattered photon yields with results from \cite{king12}, where a different method of solution was used (solving the modified wave equation for the electric and magnetic fields using Green's functions), and the dependency of photon yields on a variety of parameters was investigated. In this subsection, we choose constant parameters as in \cite{king12}. \new{First, a total power of $10\,\trm{PW}$ was shared equally between the probe and pump laser pulses, which in our case corresponds to $5\,\trm{PW}$ for the probe, and $2.5\,\trm{PW}$ for each of the two pump lasers, which are chosen to propagate in parallel and together form the single ``pump'' in \cite{king12}. The pump and probe collide head-on unless stated otherwise}, with all wavelengths set at $910\,\trm{nm}$, a probe focussed to a waist of $0.91\,\mu\trm{m}$ and the pump lasers focussed to $100\,\mu\trm{m}$, with all pulse durations equal to $30\,\trm{fs}$. 
\newline

\begin{figure}[h!!]
 \centering\noindent
  \includegraphics[draft=false, width=0.8\linewidth]{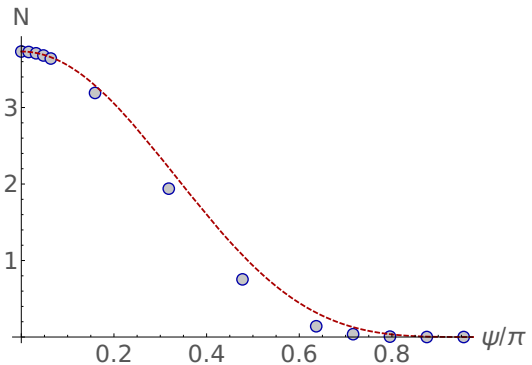}
\caption{Two beam result. Dependency of total number of photons, $N(\psi)$ on the angle, $\pi -\psi$, between propagation directions of pump and probe pulses. The dashed line is proportional to $(1+\cos\psi)^{2}$. The other parameters are as in \cite{king12}.}
\label{fig:thetaplot}
\end{figure}

Since we consider several different collision set-ups, it is important that the dependency of the rate on the collision angle is correct. Then, starting with a head-on collision and defining the collision angle $\psi=0$ at this point, we calculate how the number of scattered photons changes as we increase $\psi$. It is well-known \cite{narozhny69} that the polarisation operator predicts a change in refractive index for low frequencies and weak field strengths that is proportional to $(1+\cos\psi)^{2}$. Agreement with our prediction is shown in \figref{fig:thetaplot}.
\newline

\begin{figure}[h!!]
 \centering\noindent
  \includegraphics[draft=false, width=0.8\linewidth]{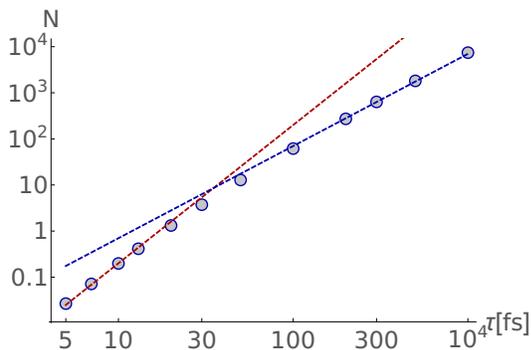}
\caption{Two beam result. Dependency of total number of photons, $N(\psi)$ on the pulse duration $\tau$, set equal for all beams. The red dashed line fit for lower $\tau$ is proportional to $\tau^{3}$, whereas the blue dashed line fit for larger $\tau$ is proportional to $\tau^{2}$. The other parameters are as in \cite{king12}.}
\label{fig:tauplot}
\end{figure}

In the XFEL set-up, we choose different pulse durations, so it should be checked that the dependency on this variable of the total number of scattered photons is as expected. In \figref{fig:tauplot} we see that for $\tau\lesssim 30\,\trm{fs}$, the dependency is approximately $\tsf{N}\propto\tau^{3}$, as in \cite{king12}, but for $\tau \gtrsim 100\,\trm{fs}$, the dependency is only $\tsf{N}\propto\tau^{2}$, compared to $\tsf{N}\propto\tau$ in \cite{king12}. The reason for the different behaviour at larger $\tau$, is because there are no decaying focal terms in our IRLA expression, and these were identified as being the new length scale, at larger $\tau$, that defined the longitudinal interaction length. So whilst we expect agreement at small $\tau$, disagreement at larger values (which are not used in any experimental scenarios studied in the main paper), is acceptable.
\newline

\begin{figure}[h!!]
 \centering\noindent
  \includegraphics[draft=false, width=0.8\linewidth]{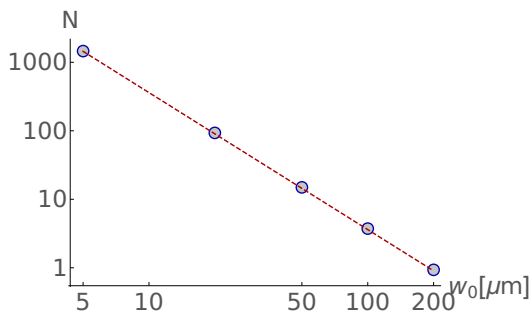}
\caption{Two beam result. Dependency of total number of photons, $\tsf{N}$ on the focal width of the probe pulse. The dashed line shows a fit of $\tsf{N}\propto w_{0}^{-2}$ as in \cite{king12}.}
\label{fig:w0plot}
\end{figure}

Another test of the behaviour comes from varying the focal width of the probe pulse, the dependency of which is plotted in \figref{fig:w0plot}. We find $\tsf{N}\propto w_{0}^{-2}$, as also reported in \cite{king12}. 

\subsection{Behaviour of scattering rates for three-pulse case}
\begin{figure}[h!!]
 \centering\noindent
  \includegraphics[draft=false, width=0.7\linewidth]{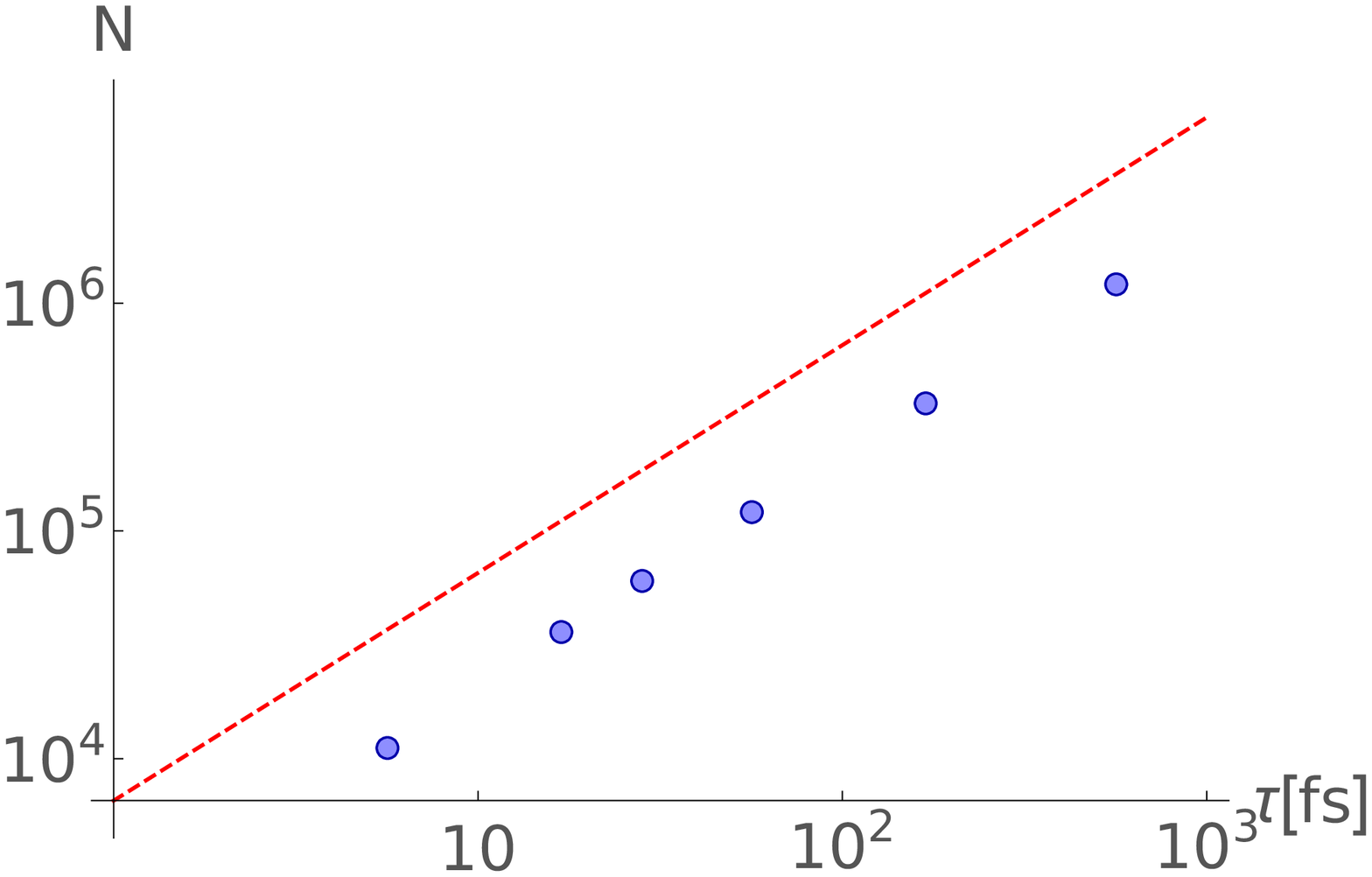}\\
    \includegraphics[draft=false, width=0.75\linewidth]{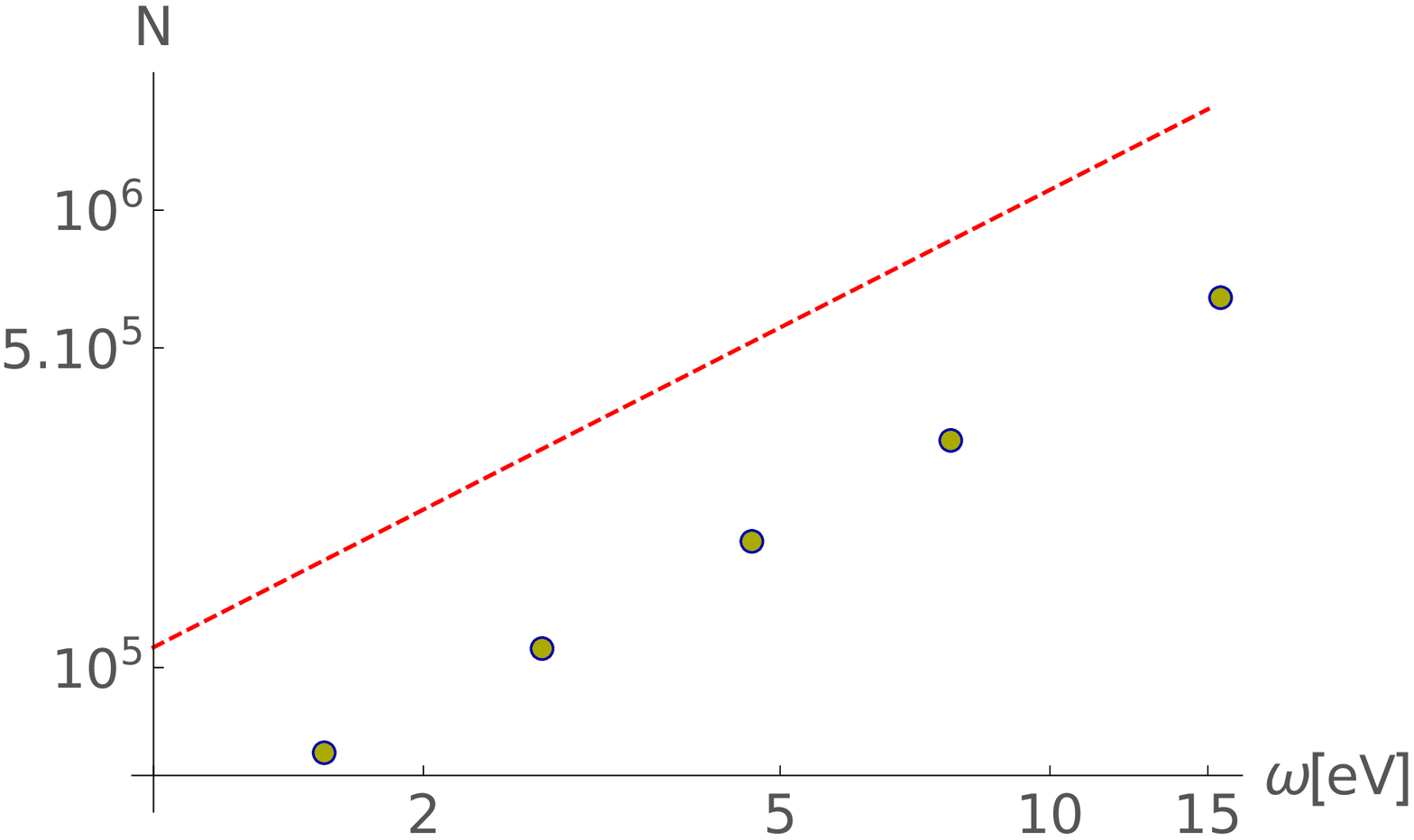}
\caption{Three beam result. The dashed line is the prediction based on the formula given in \cite{lundstroem_PRL_06}. One expects our more realistic description of the laser pulses to predict a lower number of photons than one based on a cubic interaction region of constant field strength.}
\label{fig:lstroem1}
\end{figure}

The most straightforward comparison is with the work of Lundstr\"{o}m et al. \cite{lundstroem_PRL_06}. In that work, a simple formula for predicting the number of scattered photons, for a cross-sectional beam width of $b^{2}$ for $b=1.6\mu \trm{m}$, is given as:
\[
  \tsf{N} = 0.025\,|\mbf{G}| \frac{P_{1}P_{2}P_{3}}{(1\trm{PW})^{3}}\frac{c\tau}{1\mu \trm{m}}\left(\frac{1\,\mu\trm{m}}{\lambda}\right)^{3} \left(\frac{\lambda}{800\,\trm{nm}}\right)^{2},
\]
where $|\mbf{G}| = 0.77$ \cite{lundstroemthesis}. To test this, we set $w_{0}=1.6/\sqrt{\pi}\,\mu\trm{m}$, so that the (circular) cross-sectional area of our pulses is equal to that used in \cite{lundstroem_PRL_06}. Also, because we have a Gaussian pulse envelope, which integrates to $\tau \sqrt{\pi}$, to compare this with the cubic model in \cite{lundstroem_PRL_06}, we divide our pulse duration by $\sqrt{\pi}$, so that the integral is the same. Then, unless the respective parameter is being altered, we set $P_{1}=P_{2}=P_{3}=25\,\trm{PW}$, $\tau=30\,\trm{fs}$, $\omega = 1.55\,\trm{eV}$ ($\lambda=800\,\trm{nm}$). We then find the agreement in the dependency on pulse duration and frequency in \figref{fig:lstroem1}.

\subsection{Accuracy of IRLA}
\begin{figure}[h!!]
 \centering\noindent
  \includegraphics[draft=false, width=0.45\linewidth]{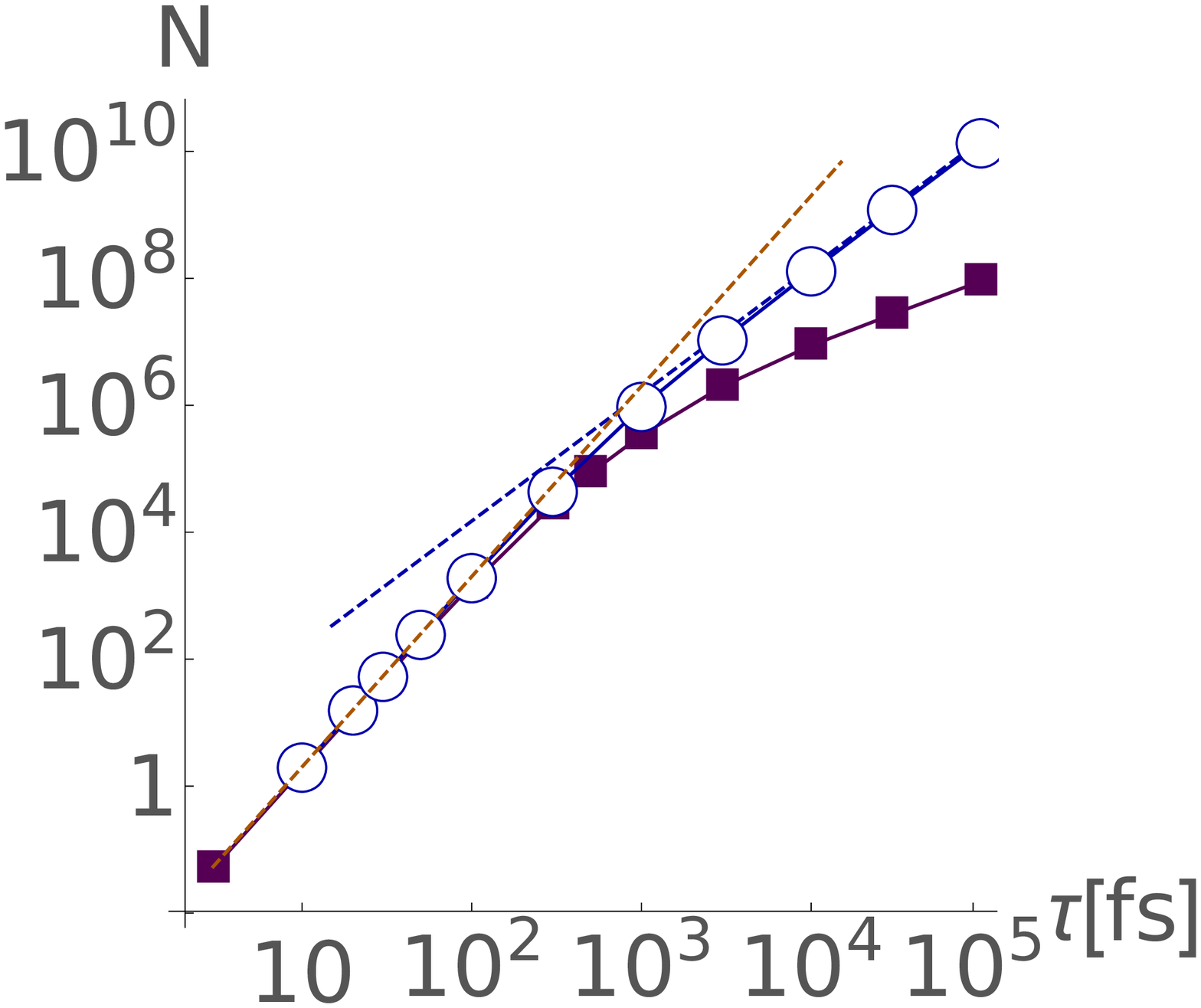}\hfill
    \includegraphics[draft=false, width=0.45\linewidth]{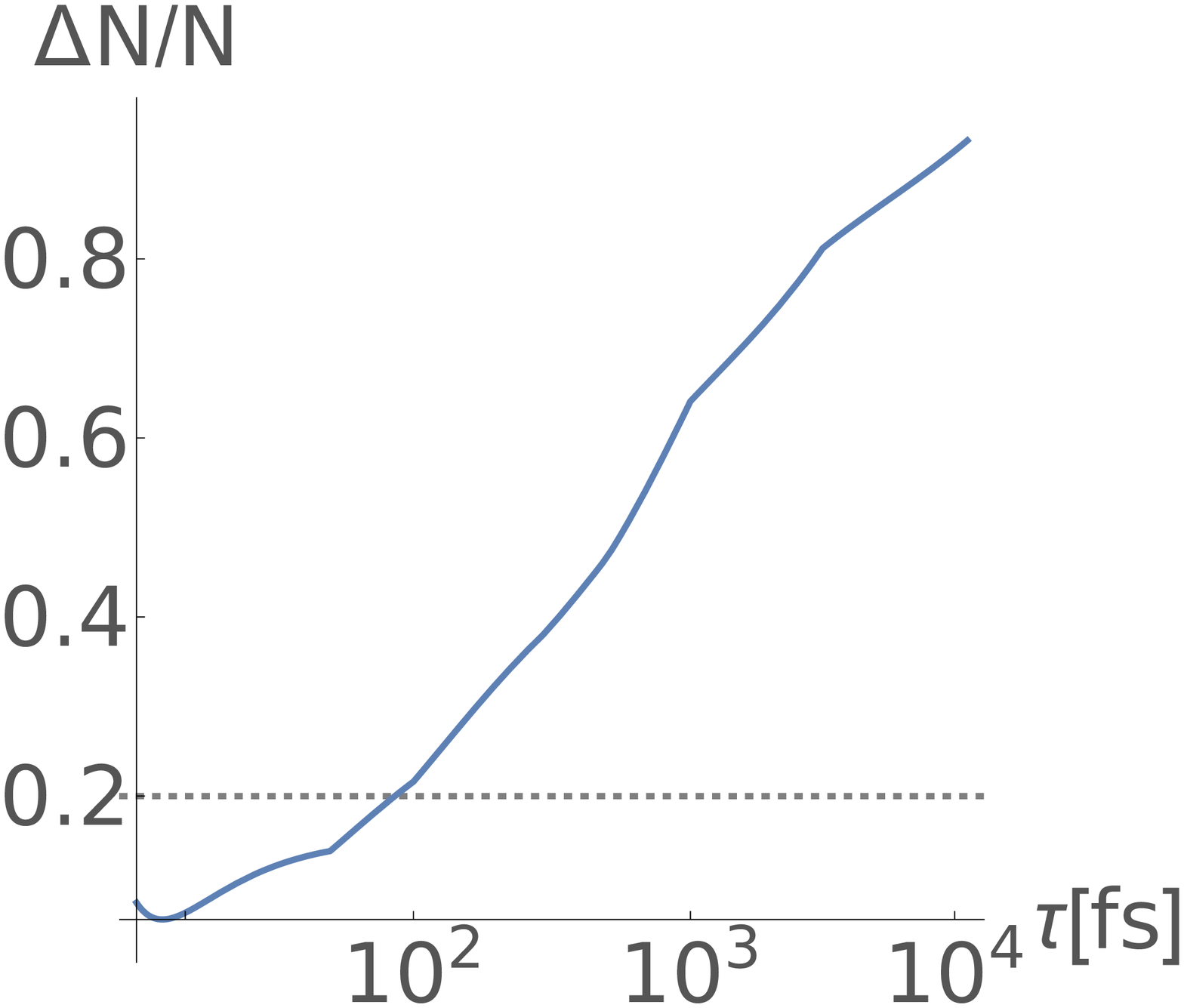}
\caption{Left: How the yield of scattered photons in the IRLA (empty circles) compares with the yield of photons from the full numerical evaluation (square markers). For $\tau \lesssim 300\,\trm{fs}$, in both cases, the yield $N\propto \tau^{3}$, whereas for larger $\tau$, the IRLA shows a dependency of $N\propto\tau^{2}$ compared to a dependency $N\propto\tau$ for the full numerical evaluation. Right: $\Delta N$ is the difference in photon yield between the IRLA and the numerical evaluation. For $\tau < 100\,\trm{fs}$, the error in the IRLA is less than $20\%$.}
\label{fig:IRLA1}
\end{figure}

In all numerical results in the main part of the paper, we have employed the Infinite Rayleigh-Length Approximation (IRLA). This removes terms which describe the curvature of wavefronts and it ignores the widening of the pulse width as one moves away from the focal spot. One expects the IRLA to lead to an \emph{over-prediction} of the number of scattered photons because the intensity of the pulse is maintained in the longitudinal direction, rather than the energy spreading out over a wider area as the pulse defocusses. Therefore, the IRLA should become progressively worse as the duration of all pulses is increased. To test how applicable the IRLA is in the parameter region of interest in the paper, we compare the predicted number of photons from the IRLA with numerically evaluating the full integral (hence calculating the spacetime integral in \eqnref{eqn:I} after performing an analytical integration in $t$ which is independent of the IRLA), for varying pulse durations $\tau$. We take pulse $1$ to have a power of $5\,\trm{PW}$ to collide head-on with the other two pulses, propagating in parallel, each with a power of $2.5\,\trm{PW}$. All pulses have a wavelength of $910\,\trm{nm}$ and are focussed to $5\,\mu\trm{m}$, which is the minimal focussing used for optical pulses in the manuscript. The results are plotted in \figref{fig:IRLA1}.
\newline

\figref{fig:IRLA1} supports our employing the IRLA in the main paper, where the maximum optical pulse duration used was $30\,\trm{fs}$. It also supports the different large-$\tau$ scaling of the photon yield reported in \figref{fig:tauplot} as being due to the absence of focussing terms.

\bibliography{current}
\end{document}